\documentclass[a4paper,12pt]{article}
\usepackage[utf8]{inputenc}
\usepackage[english]{babel}
\usepackage{authblk}
\usepackage{graphicx}
\usepackage{mathptmx}
\usepackage[singlespacing]{setspace}
\usepackage[headheight=1in,margin=1in]{geometry}
\usepackage{fancyhdr}
\usepackage{biblatex} 
\usepackage{hyperref}
\usepackage{csquotes}

\makeatletter
\def\@maketitle{%
  \newpage

  \begin{center}%
  \let \footnote \thanks
    {\LARGE \@title \par}%
  \end{center}%
  \par
  \vskip 0.1em}
\makeatother

\graphicspath{{images/}}

\title{Same View, Different Visions - Visual Nature Content on Social Media}

\date{}

\begin{document}

\maketitle
\thispagestyle{fancy}

\begin{center}
Oliver Mel Allen, Autonomous University of Barcelona \\
Johannes Langemeyer, Autonomous University of Barcelona \\
Ramin Soleymani-Fard, Barcelona Supercomputing Center
\end{center}
Real-world human–nature connections are declining rapidly, with potentially severe consequences for how societies value and protect nature. At the same time, new — and widely overlooked — forms of nature experiences are emerging in virtual spaces. While digital nature values might constitute novel incentives for environmental stewardship, they might also be appropriated by those who seek to undermine sustainable transformation. \par
As of April 2025, 88.4\% of online adults worldwide say that they use social media at least once a week, and on average they spend over 7 hours per week on social media \cite{kemp_simon_digital_2025}. The rise in digitization has fostered the loss of direct human-nature interactions and an “extinction of experience” \cite{miller_biodiversity_2005}. At the same time values generated from human relationships with nature are the foundation for global, transformative change \cite{chan_editorial_2018, obrien_ipbes_2025} and social media is a possible arena for the creation of these values. For example, growing a tomato in a community garden includes notions of care, social relations, and reciprocity that foster global environmental stewardship and transitions toward global sustainability. If one then posts a photo of this tomato on Facebook, the values of care, social relations, and reciprocity are shared with others who may not have direct access to this experience. \par
On social media, where images and videos dominate \cite{kemp_simon_digital_2025}, visual representations of nature are particularly relevant to these 'digital nature values' \cite{calcagni_plural_2023}. While qualitative studies on TikTok and Youtube show that visual aspects and framing of nature-related issues on these platforms are salient, the literature is still missing large-scale analysis focused on visual communication \cite{hautea_showing_2021, yin_tiktoks_2023, olausson_making_2020, morner_hunting_2017}. One promising method is the use of VLLMs to bolster image analysis, for example Arminio et al. \cite{arminio_leveraging_2025} used textual descriptions generated with VLLMs to cluster images related to climate debate and found these clusters were of higher quality, more interpretable, and better suited to social science research than those generated by previous computer vision approaches. \par
Digital nature values formed through online interactions with nature might incentivize new types of environmental stewardship \cite{langemeyer_virtual_2022}, but these values may also be appropriated by those who seek to undermine sustainable transformation, such as right-wing populist movements and fossil fuel industries \cite{peng_automated_2023, noauthor_facts_nodate, hughes_ecofascism_2022}. For instance, rightwing populist movements promote “back to nature” values and localization and actively disregard the Global South \cite{noauthor_facts_nodate, hughes_ecofascism_2022}, and naturalistic scenes edited to include white supremacist symbols are a hallmark of online eco-fascism \cite{hughes_ecofascism_2022}. Specific aesthetics and visual cultures may be appropriated, borrowing their credibility or influence \cite{bastos_guy_2023}, and layers of platform affordances, internet cultures, and norms add additional complexities to studies of visual communication of human-nature relationships on social media \cite{hautea_showing_2021}. For example, disinformation accounts employed by the Internet Research Agency tailored their propaganda using profile images attuned to subcultural and visual affordances of social platforms \cite{bastos_guy_2023}. Research on this phenomenon is especially important as generative AI tools for photos and video are made readily available to social media users \cite{allyn_sora_2025} and far-right political ecology is rapidly evolving \cite{allen_political_2024}. \par
\textbf{In this ongoing work we ask the following research question: What is the current state of digital human nature relationships on social media?} More specifically, we will explore  how aesthetics and visual value framings are created and spread by segments of society online. Hereby, we follow the hypothesis that nature aesthetics are increasingly appropriated to uphold the status quo and potentially push far-right ideologies countering transformative change. \par

\subsubsection*{Data and Methods}
\begin{figure}[htp]
\centering
\includegraphics[width=0.9\textwidth]{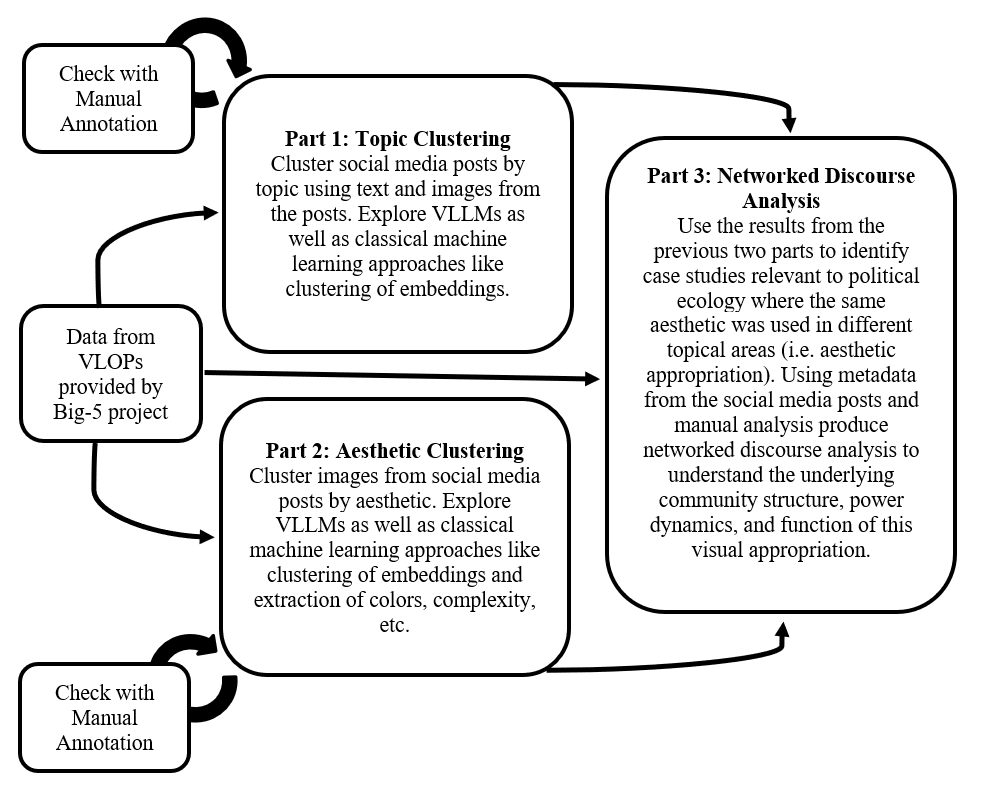}
\caption{Data and methods for the proposed study.}
\label{fig:image}
\end{figure}

The study includes three parts (see Figure \ref{fig:image}) which strategically combine natural language processing, image clustering, and networked discourse analysis to illuminate the intersection of online aesthetic appropriation, nature values, and political ecology. Data will be provided through the \href{big-5.eu/}{Big-5 ERC project}, which provides access to samples from various very large online platforms (VLOPs). \par
In parts 1 and 2 we will apply unsupervised machine learning techniques to cluster social media posts by topic, including both image and text data, (part 1) and by aesthetics (part 2). In this case we use the term \textit{digital nature aesthetics} to refer to visual signals that are connected to a specific digital culture, movement, etc., rather than referring to beauty or quality of an image. Computer vision can distinguish human-nature interactions from photographs \cite{vaisanen_exploring_2021}, and feature extraction with neural networks, Google Cloud API and Clarifai paired with clustering analysis is a common method to extract patterns from large nature image datasets gathered from social media sites like Flickr \cite{egarter_vigl_harnessing_2021, lee_mapping_2019, huai_using_2022, vaisanen_exploring_2021, song_using_2020}. However, these applications mainly focus on identification of human-nature interactions, which may not be nuanced enough for studies of visual communication and aesthetic appropriation, and further research is necessary to establish the optimal approach for validating image clustering results \cite{peng_automated_2023}. Multimodal Large Language Models (MLLMs), also known as Visual and Large Language Models (VLLMs), offer a promising new frontier for studying social media images, moving beyond object detection towards deep descriptions akin to human annotations \cite{arminio_leveraging_2025}. \par
Due to the potential for bias in automated approaches and the complexity of human-nature relationships on social media, in part 3 we will bolster our large-scale computational findings with case studies and small scale manual analysis, adapting previous approaches (e.g. \cite{lenda_effects_2020, boulianne_engagement_2023, chen_how_2023, allen_flowers_2026}). We will employ networked discourse analysis using the metadata available from social media posts in combination with manual annotation to construct a case study of aesthetic appropriation. This mixed-methods, cross-platform approach should allow us to disentangle layers of platform affordances, internet cultures, and norms. \par

\subsubsection*{Conclusion} 
Through a careful combination of unsupervised clustering, networked discourse analysis, and manual coding, this study is expected to continue to illuminate the intersection between social media, far-right and extremist groups, and human-nature relationships while advancing the theory of digital nature values and visual appropriation from a critical perspective. The study will showcase that posting hiking photos and selfies with your pet is not only a fun pastime, but an important tool to cultivate human relationships with nature that has the potential to inspire or hamper transformative change. \par

\newpage

\printbibliography

\newpage

\end{document}